\documentclass[leqno]{amsart}

\usepackage[foot]{amsaddr}

\usepackage{color}
\usepackage{geometry}
\usepackage{subfiles}
\usepackage{graphicx}
\usepackage[section]{placeins}
\usepackage{hyperref}
\usepackage{stmaryrd}
\usepackage{amsmath,amssymb,amsfonts,amsthm,textcomp}
\usepackage{mathtools}
\usepackage{tensor}
\usepackage{multicol}
\usepackage{subfig}
\usepackage{enumerate}

\newfont{\tenbfsl}{cmbxti9 scaled 1200}
\newfont{\tenbbb}{msbm10}
\newfont{\svnbbb}{msbm8}

\newcommand{\bs}[1]{\boldsymbol{#1}}
\newcommand{\br}[1]{\boldsymbol{\mathrm{{#1}}}}

\newcommand{\cl}[1]{\mathcal{#1}}

\newcommand{\id}{\bs{1}}

\newcommand{\surp}[1]{\{\!\!\{ {#1} \}\!\!\}}

\newcommand{\fr}[2]{{\textstyle{\frac{{#1}}{{#2}}}}}

\newcommand{\dv}{\,\mathrm{d}v}
\newcommand{\da}{\,\mathrm{d}a}
\newcommand{\ds}{\,\mathrm{d}\sigma}

\newcommand{\bdy}{\cl{B}}
\newcommand{\dbdy}{\partial\cl{B}}
\newcommand{\prt}{\cl{P}}
\newcommand{\dprt}{\partial\cl{P}}
\newcommand{\srf}{\cl{S}}
\newcommand{\dsrf}{\partial\cl{S}}

\newcommand{\edg}{\cl{C}}

\newcommand{\trans}{\scriptscriptstyle\mskip-1mu\top\mskip-2mu}

\newcommand{\sperp}{\scriptscriptstyle\mskip-2mu\perp\mskip-2mu}

\newcommand{\tr}{\mathrm{tr}\mskip2mu}
\newcommand{\sym}{\mathrm{sym}\mskip2mu}
\newcommand{\skw}{\mathrm{skw}\mskip2mu}
\newcommand{\ax}{\mathrm{ax}\mskip2mu}

\newcommand{\Grad}{\mathrm{grad}\mskip2mu}
\newcommand{\Div}{\mathrm{div}\mskip2mu}
\newcommand{\Curl}{\mathrm{curl}\mskip2mu}
\newcommand{\Grads}{\Grad_{\mskip-2mu\scriptscriptstyle\cl{S}}}
\newcommand{\Divs}{\Div_{\mskip-6mu\scriptscriptstyle\cl{S}}}

\newcommand{\intvspace}{\vphantom{\int\limits_{\dprt\dbdy_\tau}}}

\theoremstyle{remark}

\theoremstyle{definition}

\newcounter{syn}[section] 
\renewcommand{\thesyn}{\arabic{section}.\arabic{syn}}

\newcommand{\twovdots}{\mskip+2mu\colon\mskip-2mu}
\makeatletter
\def\threevdots{\mskip+4mu\vbox{\baselineskip2.25\p@ \lineskiplimit\z@
  \kern4.9\p@\hbox{.}\hbox{.}\hbox{.}}\mskip+3.8mu}
\makeatother

\newcommand{\bsts}{\bs{t}_{\scriptscriptstyle\cl{S}}}
\newcommand{\bstss}{\tensor*[^2]{\bs{t}}{_{\scriptscriptstyle\cl{S}}}}

\newcommand{\bstds}{\bs{t}_{\scriptscriptstyle\partial\cl{S}}}
\newcommand{\bstc}{\bs{t}_{\scriptscriptstyle\cl{C}}}

\newcommand{\bsms}{\bs{m}_{\scriptscriptstyle\cl{S}}}

\newcommand{\bstsast}{\bs{t}_{\scriptscriptstyle\cl{S}^{\ast}}}

\newcommand{\bsmsast}{\bs{m}_{\scriptscriptstyle\cl{S}^{\ast}}}

\begin{document}

\title{On the control volume arbitrariness in the Navier--Stokes equation}
\author{Luis Espath}
\address{Department of Mathematics, RWTH Aachen University, Pontdriesch 14-16, 52062 Aachen, Germany.}
\email{espath@gmail.com}

\date{\today}

\begin{abstract}
\noindent
We present a continuum theory to demonstrate the implications of considering general tractions developed on arbitrary control volumes where the surface enclosing it lacks smoothness. We then tailor these tractions to recover the Navier--Stokes-$\alpha\beta$ equation and its thermodynamics. Consistent with the surface balances postulated to propose this theory, we provide an alternative approach to derive the natural boundary conditions.
\\
\textbf{AMS subject classifications:}
$\cdot$
76A02 
$\cdot$
80A17 
$\cdot$
35L65 
$\cdot$

\end{abstract}

\maketitle

\tableofcontents                        


\section{Introduction}

Holm et al. \cite{Hol98a,Hol98b} introduced the Lagrangian averaged Euler equation. It was soon generalized by Chen et al. \cite{Che98,Che99a,Che99b} to account for viscous effects, yielding the Lagrangian averaged Navier--Stokes-$\alpha$ equation. The Navier--Stokes-$\alpha$ equation models statistically homogeneous and isotropic turbulent flows in terms of the filtered velocity. Fried \& Gurtin \cite{Fri07,Fri08} derive the Navier--Stokes-$\alpha$ and Navier--Stokes-$\alpha\beta$ continuum theories within the virtual power framework proposed by Gurtin \cite{Gur02} and Fried \& Gurtin \cite{Fri06}. Here, one should bear in mind that the continuum framework by Fried \& Gurtin, to some extent, generalizes the work by Toupin \cite{Tou62,Tou64}.

In this work, we adapt Fosdick's approach to deriving a more general Navier--Stokes-$\alpha\beta$ equation, its thermodynamics, and natural boundary conditions by considering control volumes that lack smoothness on their surface boundaries. First, recall that Fosdick \& Virga \cite{Fos89} provided a variational proof of the stress theorem of Cauchy and, using an analogous framework, Fosdick \cite{Fos16} extended his previous work and provided a variational proof of the hyperstress theorem for second-gradient theories aiming at generalizing Toupin's theory. Within Fosdick's framework while considering control volumes that lack smoothness on their surface boundaries and tailoring the tractions developed on edges, we obtain the Navier--Stokes-$\alpha\beta$ equation. Moreover, we use the surface balances postulates to obtain a broad set of natural boundary conditions.

Our notation is as follows. Linear transformations are denoted with brackets; that is, the linear transformation $\br{A}$ transforms the element $\bs{u}$ and reads $\br{A}[\bs{u}]$. The differential operators gradient, divergence, and curl are respectively denoted by $\Grad$, $\Div$, and $\Curl$. $(\cdot)^{\trans}$ is the minor transposition of the last two indices while $(\cdot)^{\sperp}$ is the minor transposition of the first two indices. Last, $\times$ and $\otimes$ are the cross and tensor products. Let a material region $\cl{A}$ undergo deformation such that $\bs{y}(\cl{A})\coloneqq\cl{A}_\tau$ represents the deformed configuration, where $\bs{y}(\bs{x},t)\in\cl{A}_\tau$ and $\bs{x}\in\cl{A}$. Through what follows, we distinguish three different regions. These regions are a material part $\cl{A}$, a spatial part $\cl{A}_\tau$, and a control volume $\overline{\cl{A}}$, and this classification applies for $\cl{A}$ being volumes, surfaces, and curves. Note that, in contrast to the boundary $\overline{\partial\cl{A}}$, material cannot migrate across $\partial\cl{A}_\tau$.

\begin{figure}[!htb]
\centering
  \begingroup
  \captionsetup[subfigure]{width=0.475\textwidth}
  \subfloat[Nonsmooth open surface $\srf$ oriented by an unit normal $\bs{n}$ with a junction-edge $\edg$ defined by the unit normals $\{\bs{n}^+,\bs{n}^-\}$ and oriented by the unit tangent $\bs{\sigma}$, and with a boundary-edge $\dsrf$ oriented by a tangent unit $\bs{\sigma}$.]{\label{fg:1}\includegraphics[width=0.475\textwidth]{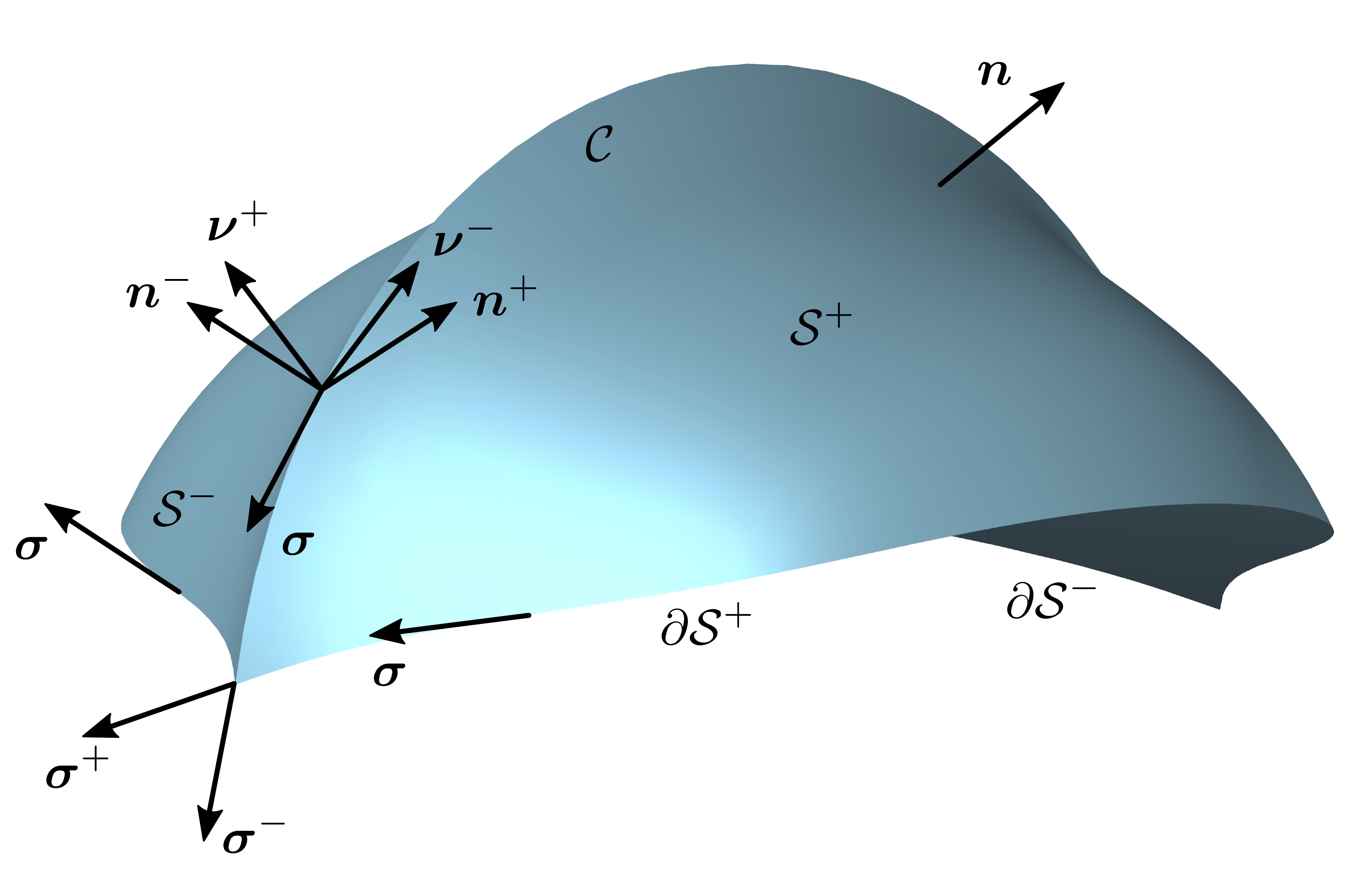}} \hspace{0.5cm}
  \subfloat[Nonsmooth part $\prt$ oriented by an unit normal $\bs{n}$ with a junction-edge $\edg$ defined by the unit normals $\{\bs{n}^+,\bs{n}^-\}$ and oriented by the unit tangent $\bs{\sigma}\coloneqq\bs{\sigma}^+$.]{\label{fg:2}\includegraphics[width=0.475\textwidth]{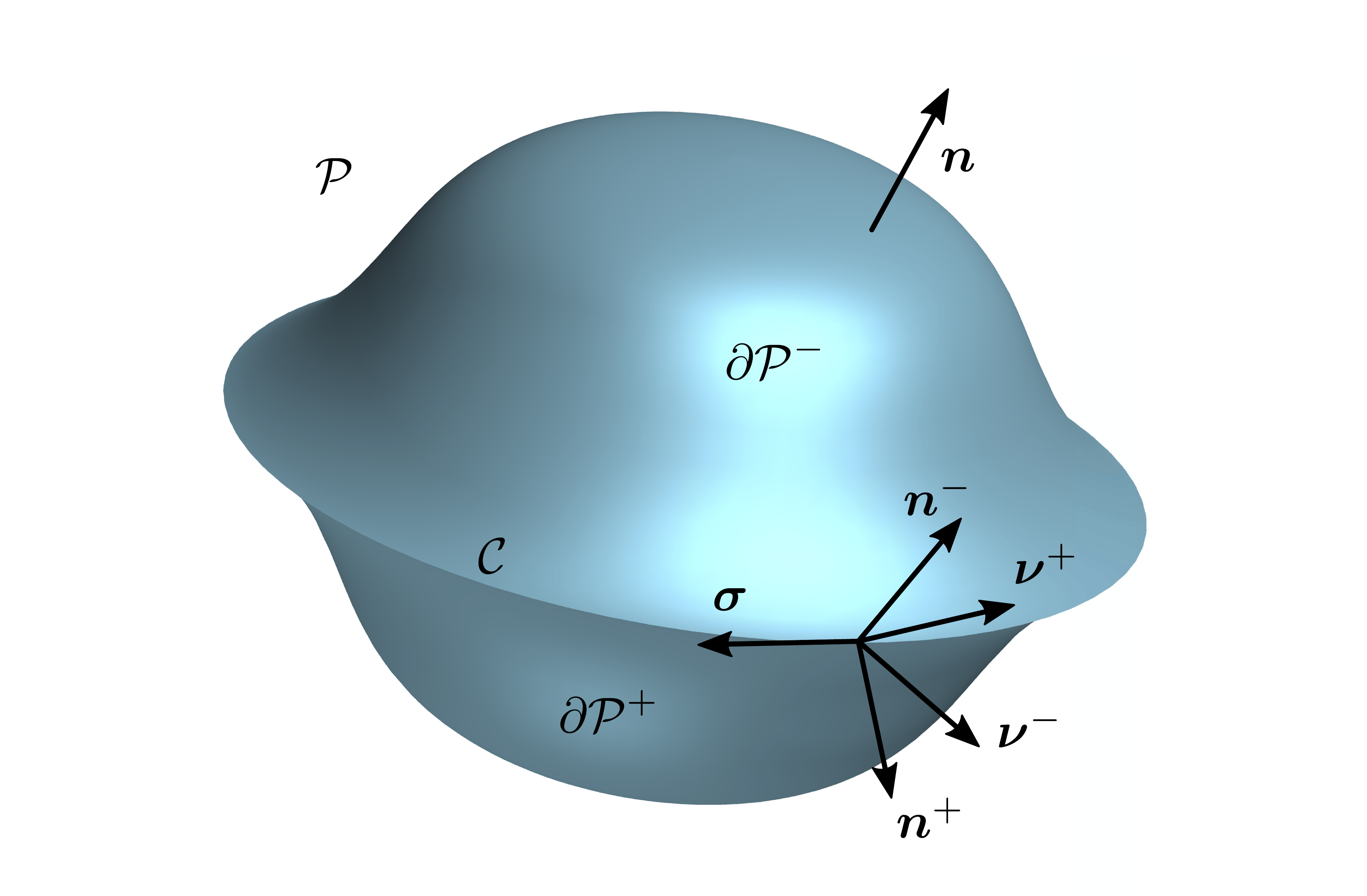}}
  \endgroup
  \caption{A nonsmooth surface (left) and a nonsmooth part (right).}
\label{fg:surfaces}
\end{figure}

Next, consider an arbitrary part $\prt$ inside a region $\bdy$ of a three-dimensional point space $\cl{E}$. To derive the traction fields, we postulate balances of forces and torques on open surfaces. A surface $\srf\subset\prt$ may lose its smoothness along a curve $\edg$. This curve represents the junction of two smooth surfaces; we then name it junction-edge. Analogously, we name its boundary $\dsrf$ as boundary-edge. A Darboux frame, $\{\bs{\sigma},\bs{n},\bs{\nu}\}$, is our choice to describe boundary- and junction-edges, where $\bs{\sigma}$ is the unit tangent, $\bs{n}$ is the unit normal, and $\bs{\nu}$ the unit tangent-normal. In considering arbitrary parts, we define a set of traction fields on each geometrical feature. That is, for the nonsmooth open surface $\srf$ in Figure \ref{fg:1}, we assume that internal interactions develop the following tractions
\begin{enumerate}[\itshape(i)]
  \item Surface traction [force/area]: $\bsts(\bs{x},t,\bs{n},\br{K})$ on $\srf$, with $\br{K}$ the curvature tensor;
  \item Surface-couple traction [torque/area]: $\bsms(\bs{x},t,\bs{n})$ on $\srf$;
  \item Boundary-edge traction [force/length]: $\bsts(\bs{x},t,\bs{n},\bs{\nu})$ on $\dsrf$;
  \item Junction-edge traction [force/length]: $\bsts(\bs{x},t,\bs{n}^{\pm})$ on $\edg$.
\end{enumerate}
Moreover, on the opposite side of $\srf$, that is, $\srf^\ast$, additional tractions are developed to counterbalance the tractions formerly presented. These are the surface traction $\bstsast$ and surface-couple traction $\bsmsast$, which are respectively the intrinsic counterparts of $\bsts$ and $\bsms$. These tractions are also developed by the contact of $\srf^\ast$ with the adjacent parts of $\bdy$.

In considering a nonsmooth part $\prt$ in Figure \ref{fg:2}, we assume that the following tractions are developed
\begin{enumerate}[\itshape(i)]
  \item Body force [force/volume]: $\bs{b}(\bs{x},t)$ on $\prt$;
  \item Surface traction [force/area]: $\bsts(\bs{x},t,\bs{n},\br{K})$ on $\dprt$;
  \item Surface-couple traction [torque/area]: $\bsms(\bs{x},t,\bs{n})$ on $\dprt$;
  \item Junction-edge traction [force/length]: $\bsts(\bs{x},t,\bs{n}^{\pm})$ on $\edg$.
\end{enumerate}

Next, we develop the explicit form of these traction fields, the field equations, thermodynamics for a purely mechanical theory, and the natural boundary conditions.

\section{Differential and integro-differential relations}

Given the coordinates $\tau^i$ $(i=1,2,3)$, the $i$th contravariant basis $\bs{g}^i$, and the conventional partial derivative $\partial_i\coloneqq\partial/\partial\tau^i$, let $\bs{\kappa}$ be a smooth vector field on $\bdy$. In this section, however, $\bs{\kappa}$ may be understood as a tensor as well. The gradient of a vector field $\bs{\kappa}$ is defined as
\begin{equation}\label{eq:gradient.vector}
\Grad\bs{\kappa}\coloneqq\partial_i\bs{\kappa}\otimes\bs{g}^i.
\end{equation}

Now, consider a smooth surface $\srf\subset\prt$ oriented by the unit normal $\bs{n}$ at $\bs{x}\in\srf$. Let $\srf$ be parameterized by coordinates $\tau^\alpha$ with $\alpha=1,2$ and $\bs{z}$ be a smooth extension of $\srf$ along its normal $\bs{n}$ at $\bs{x}$,
\begin{equation}\label{eq:parameterization.surface.extension}
\bs{z}(\bs{x},\tau)\coloneqq\bs{x}+\tau\bs{n}(\bs{x}),\qquad\forall\,\bs{x}\in\srf,
\end{equation}
with $\tau$ representing the normal coordinate $n$ and taking values in an open interval of zero so that there exists a one-to-one mapping $\bs{z}\leftrightarrow(\bs{x},\tau)$. This parameterization induces the following local covariant basis
\begin{equation}\label{eq:covariant.basis}
\bs{g}_\alpha\coloneqq\partial_\alpha\bs{z}=\partial_\alpha\bs{x}+\tau\partial_\alpha\bs{n},\qquad\bs{e}^\alpha\coloneqq\bs{g}^\alpha|_{\tau=0}.
\end{equation}
Thus, the gradient of $\bs{\kappa}$ given in expression \eqref{eq:gradient.vector} at $\bs{x}\in\srf$ takes the form
\begin{equation}\label{eq:gradient.vector.surface.extension}
\Grad\bs{\kappa}=\partial_n\bs{\kappa}\otimes\bs{n}+\partial_\alpha\bs{\kappa}\otimes\bs{e}^\alpha,
\end{equation}
where
\begin{equation}\label{eq:gradient.vector.normal.component}
\partial_n\bs{\kappa}=(\Grad\bs{\kappa})[\bs{n}].
\end{equation}
Next, let $\br{P}\coloneqq\br{P}(\bs{n})$ denote the projector onto the plane defined by $\bs{n}$, which reads
\begin{equation}\label{eq:tan.projector}
\br{P}\coloneqq\id-\bs{n}\otimes\bs{n}=\br{P}^{\trans}.
\end{equation}
In view of expressions \eqref{eq:gradient.vector.surface.extension} and \eqref{eq:tan.projector}, let the surface gradients of a vector field $\bs{\kappa}$, or a tensor field of order greater than zero, be
\begin{equation}\label{eq:surface.gradient}
\Grads\bs{\kappa}\coloneqq\partial_\alpha\bs{\kappa}\otimes\bs{e}^\alpha=(\Grad\bs{\kappa})\br{P}.
\end{equation}
Also, let the curvature tensor be defined by the negative surface gradient of the unit normal, that is,
\begin{equation}\label{eq:curvature.tensor}
\br{K}\coloneqq-\partial_\alpha\bs{n}\otimes\bs{e}^\alpha=-\Grads\bs{n}=\br{K}^{\trans},
\end{equation}
and the mean curvature by
\begin{equation}\label{eq:mean.curvature}
K\coloneqq\fr{1}{2}\tr\br{K}=-\fr{1}{2}\Divs\bs{n}.
\end{equation}

For any smooth tensor $\br{A}$ fields on a smooth closed oriented surface $\srf$, The surface divergence theorem states that
\begin{equation}\label{eq:smooth.divs.theo.open.S}
\int\limits_{\srf}\Divs(\br{A}\br{P})\da=\int\limits_{\dsrf}\br{A}[\bs{\nu}]\ds.
\end{equation}
Next, consider the following identity for a second-order tensor field $\br{A}$, or a tensor field of order greater than two,
\begin{equation}\label{eq:identity.div.surf.tensor}
\Divs\br{A}=\Divs(\br{A}\br{P})+\Divs(\br{A}[\bs{n}]\otimes\bs{n})=\Divs(\br{A}\br{P})-2K\br{A}[\bs{n}].
\end{equation}
Owing to the lack of smoothness at a junction-edge $\edg$, for any smooth tensor $\br{A}$ fields on a nonsmooth closed oriented surface $\srf$ with limiting outward unit tangent-normals $\bs{\nu}^+$ and $\bs{\nu}^-$ at $\edg$, the surface divergence theorem exhibits a \emph{surplus}, that is,
\begin{equation}\label{eq:nonsmooth.divs.theo.closed.S}
\intvspace\int\limits_{\srf}\Divs(\br{A}\br{P})\da=\int\limits_{\edg}\surp{\br{A}[\bs{\nu}]}\ds,
\end{equation}
where $\surp{\br{A}[\bs{\nu}]}\coloneqq\br{A}[\bs{\nu}^+]+\br{A}[\bs{\nu}^-]$ is the surplus term. With the identity \eqref{eq:identity.div.surf.tensor}, the surface divergence theorem \eqref{eq:nonsmooth.divs.theo.closed.S} reads
\begin{equation}\label{eq:nonsmooth.divs.theo.closed.S.tensors}
\int\limits_{\srf}\Divs(\br{A})\da=\int\limits_{\srf}2K\br{A}[\bs{n}]\da+\int\limits_{\edg}\surp{\br{A}[\bs{\nu}]}\ds.
\end{equation}
Finally, on a nonsmooth open oriented surface, the surface divergence theorem \eqref{eq:nonsmooth.divs.theo.closed.S} reads
\begin{equation}\label{eq:nonsmooth.divs.theo.open.S}
\int\limits_{\srf}\Divs(\br{A}\br{P})\da=\int\limits_{\dsrf}\br{A}[\bs{\nu}]\ds+\int\limits_{\edg}\surp{\br{A}[\bs{\nu}]}\ds.
\end{equation}
With the identity \eqref{eq:identity.div.surf.tensor}, the surface divergence theorem \eqref{eq:nonsmooth.divs.theo.open.S} reads
\begin{equation}\label{eq:nonsmooth.divs.theo.open.S.tensors}
\int\limits_{\srf}\Divs(\br{A})\da=\int\limits_{\srf}2K\br{A}[\bs{n}]\da+\int\limits_{\dsrf}\br{A}[\bs{\nu}]\ds+\int\limits_{\edg}\surp{\br{A}[\bs{\nu}]}\ds.
\end{equation}

\section{Boundary-edge, junction-edge, surface, and surface-couple tractions}

\subsection{Postulates}

We begin by postulating that the surface balance of forces
\begin{equation}\label{eq:balance.forces.S}
\int\limits_{\srf}(\bsts+\bstsast)\da+\int\limits_{\dsrf}\bstds\ds+\int\limits_{\edg}\bstc\ds=\bs{0},
\end{equation}
and the surface balance of torques
\begin{equation}\label{eq:balance.torques.S}
\int\limits_{\srf}\big(\bsms+\bsmsast+\bs{r}\times(\bsts+\bstsast)\big)\da+\int\limits_{\dsrf}\bs{r}\times\bstds\ds+\int\limits_{\edg}\bs{r}\times\bstc\ds=\bs{0},
\end{equation}
hold $\forall\,\srf\subset\bdy$ and $\forall\,t$.

Analogously, we postulate that the partwise balance of forces
\begin{equation}\label{eq:balance.forces.B}
\int\limits_{\prt}\bs{b}\dv+\int\limits_{\dprt}\bsts\da+\int\limits_{\edg}\bstc\ds=\bs{0},
\end{equation}
and the partwise balance of torques
\begin{equation}\label{eq:balance.torques.B}
\int\limits_{\prt}\bs{r}\times\bs{b}\dv+\int\limits_{\dprt}\bs{r}\times\bsts\da+\int\limits_{\edg}\bs{r}\times\bstc\ds=\bs{0},
\end{equation}
hold $\forall\,\prt\subseteq\bdy$ and $\forall\,t$ with $\bs{r}\coloneqq\bs{x}-\bs{o}$, where $\bs{o}$ is an arbitrary fixed point in $\cl{E}$.

\subsection{Boundary-edge and hyperstress specialization}

For smooth surfaces, the last integral of \eqref{eq:balance.forces.S} vanishes. Then, computing the first variation of the surface balance of forces \eqref{eq:balance.forces.S} with respect to a variation on $\srf$, Fosdick \cite[Equation (3.5)]{Fos16} arrives at the following jump condition,
\begin{equation}\label{eq:jump.edge.traction}
\llbracket{\bstds}_{\bs{\nu}}[\bs{\sigma}]\otimes\bs{\nu}+{\bstds}_{\bs{n}}[\bs{\sigma}]\otimes\bs{n}-\bstds\otimes\bs{\sigma}\rrbracket=\bs{0},
\end{equation}
when $\dsrf$ lacks of smoothness at a point, where $\llbracket(\cdot)\rrbracket\coloneqq(\cdot)^+-(\cdot)^-$ with $\pm$ consistent with `before-after' along $\bs{\sigma}$ and the subscripts in the terms ${\bstds}_{\bs{\nu}}$ and ${\bstds}_{\bs{n}}$ respectively indicate the derivative with respect to ${\bs{\nu}}$ and ${\bs{n}}$. By multiplying the jump condition \eqref{eq:jump.edge.traction} by $\bs{\sigma}^+$, one concludes that the boundary-edge traction $\bstds$ is linear with respect to $\bs{n}$ and $\bs{\nu}$, see \cite{Fos16,Esp20}. That is,
\begin{equation}\label{eq:linear.nu}
\bstds(\bs{x},t;\bs{n},\bs{\nu})\coloneqq\br{A}[\bs{\nu}],\qquad\br{A}\coloneqq\br{A}(\bs{x},t;\bs{n}),\qquad\text{and}\qquad\br{A}[\bs{n}]=0,
\end{equation}
and
\begin{equation}\label{eq:linear.n}
\bstds(\bs{x},t;\bs{n},\bs{\nu})\coloneqq\br{B}[\bs{n}],\qquad\br{B}\coloneqq\br{B}(\bs{x},t;\bs{\nu}),\qquad\text{and}\qquad\br{B}[\bs{\nu}]=0,
\end{equation}
\begin{equation*}
\forall\,\bs{n},\bs{\nu}\in\mathrm{Unit},\quad\bs{n}\cdot\bs{\nu}=0.
\end{equation*}

Aiming at encompassing conditions \eqref{eq:linear.nu} and \eqref{eq:linear.n} into a single one, we restricting attention to the cases where $\br{A}\in\mathrm{Skw}$\footnote{$\mathrm{Skw}$ is the space of all skew-symmetric transformations}. Next, we let $\{\bs{e}_1,\bs{e}_2,\bs{e}_3\coloneqq\bs{n}\}$ be an orthonormal basis, and considering the components of $\br{A}$, we have that
\begin{equation}\label{eq:skw.components.A}
\bs{e}_i\cdot\br{A}(\bs{n})[\bs{e}_\beta]=\bs{e}_\alpha\cdot\br{A}(\bs{n})[\bs{e}_\beta]\qquad\text{and}\qquad\bs{e}_\alpha\cdot\br{A}(\bs{n})[\bs{e}_\alpha]=0,
\end{equation}
where the indices $\alpha$ and $\beta$ go from $1$ to $2$, leaving out the unit normal $\bs{e}_3=\bs{n}$ from the set of orthonormal bases. Relating the components of $\br{A}$ and $\br{B}$, we have that
\begin{equation}
\bs{e}_\alpha\cdot\br{A}(\bs{n})[\bs{e}_\beta]=\bs{e}_\alpha\cdot\br{B}(\bs{e}_\beta)[\bs{n}].
\end{equation}
Thus, we can state that
\begin{align}\label{eq:boundary.edge.traction.components}
(\bs{e}_\alpha\cdot\br{A}(\bs{n})[\bs{e}_\beta])\bs{e}_\alpha\otimes\bs{e}_\beta&=(\bs{e}_\alpha\cdot\br{B}(\bs{e}_\beta)[\bs{n}])\bs{e}_\alpha\otimes\bs{e}_\beta,\nonumber\\
&=(\br{B}^{\trans}(\bs{e}_\beta)[\bs{e}_\alpha]\cdot\bs{n})\bs{e}_\alpha\otimes\bs{e}_\beta,\nonumber\\
&=(\bs{e}_\alpha\otimes\bs{e}_\beta\otimes\br{B}^{\trans}(\bs{e}_\beta)[\bs{e}_\alpha])[\bs{n}].
\end{align}

Now, noting that \eqref{eq:boundary.edge.traction.components} is $\br{A}(\bs{n})$, the boundary-edge traction $\bstds(\bs{x},t;\bs{n},\bs{\nu})=\br{A}(\bs{n})[\bs{\nu}]$ can be specified as
\begin{align}
\bstds(\bs{x},t;\bs{n},\bs{\nu})&=\big((\bs{e}_\alpha\otimes\bs{e}_\beta\otimes\br{B}^{\trans}(\bs{e}_\beta)[\bs{e}_\alpha])[\bs{n}]\big)[\bs{\nu}],\nonumber\\
&=\big((\bs{e}_\alpha\otimes\br{B}^{\trans}(\bs{e}_\beta)[\bs{e}_\alpha]\otimes\bs{e}_\beta)[\bs{\nu}]\big)[\bs{n}],\nonumber\\
&=\big(((\bs{e}_\alpha\otimes\bs{e}_\alpha)\br{B}(\bs{e}_\beta)\otimes\bs{e}_\beta)[\bs{\nu}]\big)[\bs{n}],\nonumber\\
&=\big((\br{P}\mskip-2.5mu\br{B}(\bs{e}_\beta)\otimes\bs{e}_\beta)[\bs{\nu}]\big)[\bs{n}],
\end{align}
where $\br{P}\coloneqq\bs{e}_\alpha\otimes\bs{e}_\alpha$. Expressing $\br{P}\mskip-2.5mu\br{B}(\bs{e}_\beta)\eqqcolon\overline{\br{B}}$ in a fixed orthonormal basis $\bs{e}^\prime_i$, we have that $\overline{\br{B}}(\bs{e}_\beta)=\overline{B}_{ij}^\prime(\bs{e}_\beta)\bs{e}^\prime_i\otimes\bs{e}^\prime_j$. By using this orthonormal basis, the boundary-edge traction $\bstds(\bs{x},t;\bs{n},\bs{\nu})$ assumes the form
\begin{equation}
\bstds(\bs{x},t;\bs{n},\bs{\nu})=\big((\overline{B}_{ij}^\prime(\bs{e}_\beta)\bs{e}^\prime_i\otimes\bs{e}_\beta\otimes\bs{e}^\prime_j)[\bs{n}]\big)[\bs{\nu}].
\end{equation}
Since we only consider the case where $\br{A}(\bs{n})$ is a skew-symmetric tensor, then
\begin{equation}\label{eq:skw.A}
\br{A}(\bs{n}) = \big((\overline{B}_{ij}^\prime(\bs{e}_\beta)\bs{e}^\prime_i\otimes\bs{e}_\beta\otimes\bs{e}^\prime_j)[\bs{n}]\big),
\end{equation}
also represents a skew-symmetric transformation. Moreover, a skew transformation can be expressed as an axial-vector. Therefore, there exists a linear transformation $\br{G}(\bs{x},t)\in\mathrm{Lin}$\footnote{$\mathrm{Lin}$ is the space of all linear transformations}, referred to as the `reduced' hyperstress tensor field in $\bdy$ for all $(\bs{x},t)$, such that
\begin{equation}\label{eq:boundary.edge.traction}
\bstds(\bs{x},t;\bs{n},\bs{\nu})\coloneqq\br{G}(\bs{x},t)[\bs{n}]\times\bs{\nu},
\end{equation}
where $\br{G}[\bs{n}]\times$ is the axial vector\footnote{The second-order tensor $(\bs{a}\times)$ is a `vector cross' (see the book by Gurtin et al.\ \cite[\S1]{Gur10}) and is defined such that for any vectors $\bs{a}$ and $\bs{b}$, the cross product is written as a linear transformation $(\bs{a}\times)\bs{b}=\bs{a}\times\bs{b}$. With the alternating symbol $\epsilon$, $(\bs{a}\times)=\epsilon_{ijk}a_j$.} of the skew-symmetric transformation $\br{A}$ given in \eqref{eq:skw.A}. The choice of $\br{A}$ being a skew-symmetric transformation, which renders the boundary-edge traction \eqref{eq:boundary.edge.traction}, is consistent with the Navier--Stokes-$\alpha\beta$ theory of Fried \& Gurtin \cite{Fri08}.

\subsection{Surface and surface-couple traction jumps across a surface}

In considering a smooth open oriented surface $\cl{S}$, the last integral corresponding to the junction-edge traction vanishes in the balance of forces \eqref{eq:balance.forces.S}. Replacing the boundary-edge traction \eqref{eq:boundary.edge.traction} in \eqref{eq:balance.forces.S} and applying the surface divergence theorem for smooth open surfaces \eqref{eq:smooth.divs.theo.open.S}, by localization, we arrive at
\begin{equation}\label{eq:surface.traction.opposite}
-\bstsast=\bsts+\Divs((\br{G}[\bs{n}]\times)\br{P}).
\end{equation}
Expression \eqref{eq:surface.traction.opposite} represents a jump condition across the surface. Using identity \eqref{eq:identity.div.surf.tensor}, equation \eqref{eq:surface.traction.opposite} can be expressed as
\begin{equation}\label{eq:surface.traction.opposite.alternative}
-\bstsast=\bsts+\Divs(\br{G}[\bs{n}]\times)-2K\bs{n}\times\br{G}[\bs{n}].
\end{equation}
Now, we recall the variational proof of the Cauchy stress by Fosdick \cite{Fos89} and set
\begin{equation}\label{eq:cauchy}
-\bstsast(\bs{x},t;\bs{n},\br{K})\coloneqq\br{H}(\bs{x},t)[\bs{n}],
\end{equation}
where $\br{H}(\bs{x},t)$ is a stress-like field, and its explicit form will be given later. With \eqref{eq:cauchy}, expression \eqref{eq:surface.traction.opposite.alternative} reads
\begin{align}\label{eq:surface.traction.H}
\bsts(\bs{x},t;\bs{n},\br{K})&=\br{H}[\bs{n}]-\Divs((\br{G}[\bs{n}]\times)\br{P}),\nonumber\\
&=\br{H}[\bs{n}]-\Divs(\br{G}[\bs{n}]\times)+2K\bs{n}\times\br{G}[\bs{n}].
\end{align}

Next, emulating the procedure to arrive at expression \eqref{eq:surface.traction.opposite}, we consider the balance of torques \eqref{eq:balance.torques.S} for smooth open oriented surface $\cl{S}$. Thus, the last integral, in \eqref{eq:balance.torques.S}, corresponding to the torque provoked by junction-edge traction vanishes. With the identity
\begin{equation}\label{eq:id.bsms.ast}
\bs{r}\times\mskip+1mu\Divs((\br{G}[\bs{n}]\times)\br{P})=\Divs(\bs{r}\times(\br{G}[\bs{n}]\times)\br{P})+(\br{P}-\tr(\br{P})\id)\br{G}[\bs{n}].
\end{equation}
while replacing the boundary-edge traction \eqref{eq:boundary.edge.traction} and the jump condition \eqref{eq:surface.traction.opposite} in \eqref{eq:balance.torques.S} and applying the surface divergence theorem for smooth open surfaces, we obtain the following jump condition across the surface
\begin{equation}\label{eq:surface.couple.traction.opposite}
-\bsmsast=\bsms-(\br{P}-\tr(\br{P})\id)\br{G}[\bs{n}].
\end{equation}

\subsection{Junction-edge traction}

Consider the balance of force \eqref{eq:balance.forces.S} on a nonsmooth surface with the boundary-edge traction \eqref{eq:boundary.edge.traction} and the surface traction jump condition across a surface \eqref{eq:surface.traction.opposite}. Then, applying the surface divergence theorem on nonsmooth open oriented surfaces \eqref{eq:nonsmooth.divs.theo.open.S}, by localization, we obtain the following representation for the junction-edge traction
\begin{equation}\label{eq:internal.edge.traction}
\bstc(\bs{x},t;\bs{n}^+,\bs{n}^-)=\surp{\br{G}[\bs{n}]\times\bs{\nu}},
\end{equation}
where $\surp{\br{G}[\bs{n}]\times\bs{\nu}}\coloneqq\br{G}[\bs{n}^+]\times\bs{\nu}^++\br{G}[\bs{n}^-]\times\bs{\nu}^-$.

\subsection{Field equations}

Replacing the surface traction \eqref{eq:surface.traction.H} and the junction-edge traction \eqref{eq:internal.edge.traction} into the partwise balance of forces \eqref{eq:balance.forces.B} and applying the surface divergence theorem for nonsmooth closed surfaces \eqref{eq:nonsmooth.divs.theo.closed.S}, after localization, we arrive at
\begin{equation}\label{eq:force.balance}
\bs{b}+\Div\br{H}=\bs{0}.
\end{equation}
By setting
\begin{equation}\label{eq:stress.hyperstress}
\br{T}\coloneqq\br{H}+\Div(\br{G}\times)\footnotemark,
\end{equation}
\footnotetext{$(\br{G}\times)\coloneqq\epsilon_{ikj}G_{kl}=-\epsilon_{ijk}G_{kl}$.}with the identity
\begin{equation}
\Div^2(\br{G}\times)=-\Curl\Div\br{G},
\end{equation}
the pointwise balance of forces \eqref{eq:force.balance} takes the form
\begin{equation}\label{eq:force.balance.nonclassic.form}
\bs{b}+\Div\br{T}+\Curl\Div\br{G}=\bs{0}.
\end{equation}
Using the initial and noninertial contributions of $\bs{b}=\bs{b}^{\mathrm{ni}}-\varrho\dot{\bs{\upsilon}}$, with density $\varrho$, we are led to the field equation
\begin{equation}\label{eq:force.balance.classic.form}
\varrho\dot{\bs{\upsilon}}-\Div\br{T}-\Curl\Div\br{G}-\bs{b}^{\mathrm{ni}}=\bs{0},
\end{equation}
where the dot represents the material derivative.

Next, consider the following set of identities
\begin{equation}\label{eq:id.G.1}
\Div(\bs{r}\otimes\Div(\br{G}\times))=\bs{r}\otimes\Div^2(\br{G}\times)+\Div((\br{G}\times)^{\sperp}),
\end{equation}
\begin{align}\label{eq:id.G.2}
(\br{G}[\bs{n}]\times)^{\trans}&=\left[(\br{G}[\bs{n}]\times)\br{P}+(\br{G}[\bs{n}]\times)\bs{n}\otimes\bs{n}\right]^{\trans},\nonumber\\
&=\br{P}(\br{G}[\bs{n}]\times)^{\trans}+\bs{n}\otimes(\br{G}[\bs{n}]\times)\bs{n},
\end{align}
and
\begin{equation}\label{eq:id.G.3}
\Divs(\bs{r}\otimes(\br{G}[\bs{n}]\times)\br{P})=\bs{r}\otimes\Divs((\br{G}[\bs{n}]\times)\br{P})+\br{P}(\br{G}[\bs{n}]\times)^{\trans}.
\end{equation}
Integrating \eqref{eq:id.G.1} on $\prt$, applying the volume divergence theorem, replacing the combination of \eqref{eq:id.G.2} with \eqref{eq:id.G.3}, applying the surface divergence theorem for nonsmooth closed surfaces \eqref{eq:nonsmooth.divs.theo.closed.S}, and using the definition of the axial vector\footnote{The axial vector of a second-order skew-symmetric tensor $\fr{1}{2}(\bs{a}\otimes\bs{b}-\bs{b}\otimes\bs{a})$ is given by $\fr{1}{2}\ax(\bs{a}\otimes\bs{b}-\bs{b}\otimes\bs{a})=\ax(\skw(\bs{a}\otimes\bs{b}))=-\fr{1}{2}\bs{a}\times\bs{b}$.}, we are led to the integro-differential identity
\begin{multline}\label{eq:mult.id.cross}
\intvspace\int\limits_{\prt}\bs{r}\times\Div^2(\br{G}\times)\dv=-\int\limits_{\dprt}\bs{r}\times(\bs{n}\times\Div\br{G})\da\\
\intvspace+\int\limits_{\dprt}(\bs{r}\times\Divs((\br{G}[\bs{n}]\times)\br{P})-\bs{n}\times(\br{G}[\bs{n}]\times)\bs{n})\da-\int\limits_{\edg}\bs{r}\times\surp{\br{G}[\bs{n}]\times\bs{\nu}}\ds.
\end{multline}
By replacing the surface traction \eqref{eq:surface.traction.H} and the junction-edge traction \eqref{eq:internal.edge.traction} into the partwise balance of torques \eqref{eq:balance.torques.B}, together with the identity \eqref{eq:mult.id.cross}, we have that
\begin{equation}\label{eq:balance.torque.mod.1}
\int\limits_{\prt}\bs{r}\times\bs{b}\dv-\int\limits_{\prt}\bs{r}\times\Div^2(\br{G}\times)\dv+\int\limits_{\dprt}(\bs{r}\times(\br{H}\bs{n}-\bs{n}\times\Div\br{G})+\bsms-\bs{n}\times(\br{G}[\bs{n}]\times)\bs{n})\da=\bs{0}.
\end{equation}
Now, consider the following identity obtained by applying the surface divergence theorem
\begin{equation}\label{eq:ax.id}
\int\limits_{\dprt}\bs{r}\times\br{T}\bs{n}\da=\int\limits_{\prt}(\ax(\br{T}-\br{T}^{\trans})+\bs{r}\times\Div\br{T})\dv.
\end{equation}
Last, with the definition \eqref{eq:stress.hyperstress}, the identity \eqref{eq:ax.id}, the pointwise balance of forces \eqref{eq:force.balance} in expression \eqref{eq:balance.torque.mod.1}, by localization, we obtain the following condition
\begin{equation}\label{eq:torque.balance}
\ax(\br{T}-\br{T}^{\trans})=\bs{0},
\end{equation}
which implies $\br{T}$ is a symmetry tensor. Using the $\epsilon$--$\delta$ identity\footnote{$\epsilon_{ijk}\epsilon_{imn}=\delta_{jm}\delta_{kn}-\delta_{jn}\delta_{km}$}, in the last term of the last integral of \eqref{eq:balance.torque.mod.1}, we have that
\begin{equation}\label{eq:id.n.P.G}
\bs{n}\times(\br{G}[\bs{n}]\times)\bs{n}=\br{P}\br{G}[\bs{n}].
\end{equation}
and arrive at the explicit form of the surface-couple traction
\begin{equation}\label{eq:surface.couple.traction}
\bsms=\br{P}\br{G}[\bs{n}],
\end{equation}
and substituting \eqref{eq:surface.couple.traction} into \eqref{eq:surface.couple.traction.opposite}, we have that
\begin{equation}\label{eq:surface.couple.traction.opposite.G}
-\bsmsast=\tr(\br{P})\br{G}[\bs{n}]=2\br{G}[\bs{n}].
\end{equation}
Last, with definition \eqref{eq:stress.hyperstress}, the surface traction takes the form
\begin{align}\label{eq:surface.traction}
\bsts(\bs{x},t;\bs{n},\br{K})&=\br{T}[\bs{n}]-(\Div(\br{G}\times))[\bs{n}]-\Divs((\br{G}[\bs{n}]\times)\br{P}),\nonumber\\
&=\br{T}[\bs{n}]-\Divs(\br{G}[\bs{n}]\times)+\bs{n}\times(\Div\br{G}+2K\br{G}[\bs{n}]),
\end{align}
and substituting \eqref{eq:stress.hyperstress} in \eqref{eq:surface.traction.opposite.alternative}, we obtain
\begin{equation}\label{eq:cauchy.explicit}
-\bstsast(\bs{x},t;\bs{n},\br{K})=\br{T}[\bs{n}]-(\Div(\br{G}\times))[\bs{n}].
\end{equation}

\section{Thermodynamics}

\subsection{Power balance}

In view of the field equations \eqref{eq:force.balance.nonclassic.form} and \eqref{eq:torque.balance}, in balancing the internal power expenditure
\begin{align}
\cl{W}_{\mathrm{int}}(\overline{\prt})&\coloneqq\int\limits_{\overline{\prt}}\big(\br{T}\twovdots\Grad\bs{\upsilon}+(\br{G}\times)\threevdots\Grad^2\bs{\upsilon}\big)\dv,\nonumber\\
&=\int\limits_{\overline{\prt}}\big(\br{T}\twovdots\Grad\bs{\upsilon}+\br{G}\twovdots\Grad\Curl\bs{\upsilon}\footnotemark\big)\dv,
\end{align}
\footnotetext{$\Grad\Curl\bs{\upsilon}\coloneqq\epsilon_{iab}\partial_j\partial_a\upsilon_b$; $(\br{G}\times)\threevdots\Grad^2\bs{\upsilon}=\epsilon_{imj}G_{mk}\partial_k\partial_j\upsilon_i=G_{mk}\partial_k\epsilon_{mji}\partial_j\upsilon_i=\br{G}\twovdots\Grad\Curl\bs{\upsilon}$.}with the external power expenditure
\begin{equation}
\cl{W}_{\mathrm{ext}}(\overline{\prt})\coloneqq\int\limits_{\overline{\prt}}\bs{b}\cdot\bs{\upsilon}\dv+\int\limits_{\overline{\srf}}(\bsts\cdot\bs{\upsilon}+\bstss\cdot\partial_n\bs{\upsilon})\da+\int\limits_{\overline{\edg}}\bstc\cdot\bs{\upsilon}\ds,
\end{equation}
on a control volume $\overline{\prt}$, where the surface traction $\bsts$ and the junction-edge traction $\bstc$ are respectively given in \eqref{eq:surface.traction} and \eqref{eq:internal.edge.traction}, we obtain the explicit form of the surface hypertraction $\bstss$, that is,
\begin{equation}\label{eq:hypertraction}
\bstss=\br{G}[\bs{n}]\times\bs{n}.
\end{equation}
Moreover, with identity \eqref{eq:id.n.P.G}, the surface-couple traction \eqref{eq:surface.couple.traction} can be written in term of the hypertraction \eqref{eq:hypertraction}
\begin{equation}\label{eq:hypertraction.surface.couple}
\bsms=\bs{n}\times\bstss.
\end{equation}

\subsection{Free-energy imbalance}

Restricting attention to a purely mechanical theory for incompressible materials upon the requirement that the temporal increase in the total free-energy of an arbitrary spatial region $\prt_\tau$ that advects with the body be less than or equal to the external power, we have that
\begin{equation}\label{eq:partwise.free.energy.imbalance}
\dot{\overline{\int\limits_{\prt_\tau}\varrho\psi\dv}}\le\cl{W}_{\mathrm{ext}}(\prt_\tau).
\end{equation}
Owing to the balance of mass, we have that
\begin{equation}\label{eq:partwise.mass.balance}
\dot{\overline{\int\limits_{\prt_\tau}\varrho\psi\dv}}=\int\limits_{\prt_\tau}\varrho\dot{\psi}\dv.
\end{equation}
Next, considering that $\cl{W}_{\mathrm{ext}}(\prt_\tau)=\cl{W}_{\mathrm{int}}(\prt_\tau)$, defining $\bs{\varpi}\coloneqq\Curl\bs{\upsilon}$, and with the balance of torques \eqref{eq:torque.balance} in expression \eqref{eq:partwise.free.energy.imbalance}, we arrive at the following pointwise free-energy imbalance
\begin{equation}\label{eq:pointwise.free.energy.imbalance}
\varrho\dot{\psi}\le\br{T}\twovdots\Grad\bs{\upsilon}+\br{G}\twovdots\Grad\bs{\varpi}.
\end{equation}
Last, note that $\tr(\Grad\Curl\bs{\upsilon})=\Div\Curl\bs{\upsilon}=0$; thus, we can set $\tr\br{G}$ without loss of generality.

\section{Constitutive relations}

Through what follows, we restrict attention to incompressible fluids. Here, let
\begin{equation}
\br{L}\coloneqq\Grad\bs{\upsilon},\qquad\br{D}\coloneqq\fr{1}{2}(\br{L}+\br{L}^{\trans}),\qquad\br{W}\coloneqq\fr{1}{2}(\br{L}-\br{L}^{\trans})\qquad\text{and}\qquad\br{J}\coloneqq\Grad\bs{\varpi}.
\end{equation}
For incompressible fluids, we have that
\begin{equation}
\varrho=\text{constant}\qquad\text{and}\qquad\Div\bs{\upsilon}=0.
\end{equation}
Thus, without loss of generality, the stress tensor $\br{T}$ can be decomposed into
\begin{equation}\label{eq:traceless.stress.pressure}
\br{T}\coloneqq\br{S}-p\id,\qquad\tr\br{S}=0.
\end{equation}
Accounting for the balance of torques \eqref{eq:torque.balance}, the free-energy imbalance \eqref{eq:pointwise.free.energy.imbalance} may be written as
\begin{equation}\label{eq:pointwise.free.energy.imbalance.mod}
\varrho\dot{\psi}-\br{S}\twovdots\br{D}-\br{G}\twovdots\br{J}\le0.
\end{equation}

\subsection{Constitutive hypostheses}

Within this continuum framework, let the constitutive processes $\psi$, $\br{T}$, and $\br{G}$ be given by the following constitutive response functions
\begin{equation}
\left\{\,
\begin{aligned}
&\psi=\hat{\psi}(\mathring{\br{D}},\br{D},\br{J}),\\
&\br{T}=\hat{\br{T}}(\mathring{\br{D}},\br{D},\br{J}),\\
&\br{G}=\hat{\br{G}}(\mathring{\br{D}},\br{D},\br{J}),
\end{aligned}
\right.
\end{equation}
where $\mathring{\br{D}}\coloneqq\dot{\br{D}}+\br{D}\br{W}-\br{W}\br{D}$ is the Jaumann rate of the stretch tensor $\br{D}$.

Next, restricting attention to linear constitutive response functions, consistent with the free-energy imbalance \eqref{eq:pointwise.free.energy.imbalance.mod}, by applying the Coleman--Noll procedure \cite{Col63}, we have that the constitutive response function $\psi$ only depends on $\br{D}$ and that
\begin{equation}\label{eq:constitutive.equations}
\left\{\,
\begin{aligned}
&\psi=\alpha^2|\br{D}|^2,\\
&\br{S}=2(\mu\br{D}+\varrho\alpha^2\mathring{\br{D}}),\\
&\br{G}=\mu\beta^2(\Grad\bs{\varpi}+\gamma(\Grad\bs{\varpi})^{\trans}).
\end{aligned}
\right.
\end{equation}
We refer the interested reader to the work of Fried \& Gurtin \cite{Fri08} for the physical meaning of the moduli $\mu$, $\alpha$, $\beta$, and $\gamma$. These choices \eqref{eq:constitutive.equations}, in the pointwise free-energy imbalance \eqref{eq:pointwise.free.energy.imbalance.mod}, render the following dissipation inequality
\begin{equation}
2\mu|\br{D}|^2+\mu\beta^2(1+\gamma)\mskip3mu\sym(\Grad\bs{\varpi})+\mu\beta^2(1-\gamma)\mskip3mu\skw(\Grad\bs{\varpi})\ge0,
\end{equation}
where sym and skw represent the symmetry and skew-symmetric operators. Last, considering \eqref{eq:constitutive.equations} in the balance of forces \eqref{eq:force.balance.classic.form} and the stress representation \eqref{eq:traceless.stress.pressure}, we are led to
\begin{equation}\label{eq:force.balance.classic.form.final}
\varrho\dot{\bs{\upsilon}}+\Grad{p}-\mu(1-\beta^2\Delta)\Delta\bs{\upsilon}-2\varrho\alpha^2\Div\mathring{\br{D}}-\bs{b}^{\mathrm{ni}}=\bs{0}.
\end{equation}
The Navier--Stokes-$\alpha\beta$ equation \eqref{eq:force.balance.classic.form.final} was originally obtained by Fried \& Gurtin \cite{Fri08}.

\section{Natural boundary conditions}

Here, we present a novel yet simple approach to determine boundary conditions, which is analogous to the one used by Espath \& Calo \cite{Esp20}, and it differs from the one used by Fried \& Gurtin \cite{Fri06,Fri08}. We rely on the balances of forces and torques on nonsmooth open surfaces, that is, the postulates \eqref{eq:balance.forces.S} and \eqref{eq:balance.torques.S} used to derive this continuum theory. To this end, we take the open surface $\srf$ to the limit such that the surface coincides with a portion of the boundary $\dbdy$, that is, $\srf\subseteq\dbdy$ is in contact with the environment. Thereby, the balances \eqref{eq:balance.forces.S} and \eqref{eq:balance.torques.S} specialize to
\begin{equation}\label{eq:balance.forces.S.env}
\int\limits_{\srf}(\bsts^{\mathrm{env}}+\bstsast)\da+\int\limits_{\dsrf}\bstds^{\mathrm{env}}\ds+\int\limits_{\edg}\bstc^{\mathrm{env}}\ds=\bs{0},
\end{equation}
and
\begin{equation}\label{eq:balance.torques.S.env}
\int\limits_{\srf}\big(\bsms^{\mathrm{env}}+\bsmsast+\bs{r}\times(\bsts^{\mathrm{env}}+\bstsast)\big)\da+\int\limits_{\dsrf}\bs{r}\times\bstds^{\mathrm{env}}\ds+\int\limits_{\edg}\bs{r}\times\bstc^{\mathrm{env}}\ds=\bs{0},
\end{equation}
where $\bsts^{\mathrm{env}}$, $\bstds^{\mathrm{env}}$, $\bstc^{\mathrm{env}}$, and $\bsms^{\mathrm{env}}$ are respectively the surface, boundary-edge, junction-edge, and surface-couple environmental tractions. These environmental tractions represent the external action on the boundary.

Using identity \eqref{eq:id.bsms.ast}, the surface-couple traction on the opposite side of $\srf$ in \eqref{eq:surface.couple.traction.opposite.G} can be written as
\begin{equation}\label{eq:eq:surface.couple.traction.opposite.G.id.bsms.ast}
\bsmsast=\bs{r}\times\mskip+1mu\Divs((\br{G}[\bs{n}]\times)\br{P})-\Divs(\bs{r}\times(\br{G}[\bs{n}]\times)\br{P})-\br{P}\br{G}[\bs{n}].
\end{equation}
Replacing \eqref{eq:eq:surface.couple.traction.opposite.G.id.bsms.ast} into \eqref{eq:balance.torques.S.env} and using the surface divergence for nonsmooth open oriented surface \eqref{eq:nonsmooth.divs.theo.open.S.tensors}, we arrive at the following natural boundary conditions
\begin{equation}\label{eq:traction.env}
\left\{\,
\begin{aligned}
\bsts^{\mathrm{env}}&=\br{T}[\bs{n}]-\Divs(\br{G}[\bs{n}]\times)+\bs{n}\times(\Div\br{G}+2K\br{G}[\bs{n}]),\qquad\text{on }\srf\subseteq\dbdy,\\
\bsms^{\mathrm{env}}&=\bs{n}\times(\br{G}[\bs{n}]\times\bs{n}),\qquad\text{on }\srf\subseteq\dbdy,\\
\bstds^{\mathrm{env}}&=\br{G}[\bs{n}]\times\bs{\nu},\qquad\text{on }\dsrf,\\
\bstc^{\mathrm{env}}&=\surp{\br{G}[\bs{n}]\times\bs{\nu}}\qquad\text{on }\edg.
\end{aligned}
\right.
\end{equation}

\section{Conclusion}

This continuum theory provides a derivation for the Navier--Stokes equation, where we obtain a more general representation of the now traditional Navier--Stokes-$\alpha\beta$ equation by considering control volumes that lack smoothness on their surface boundaries. We account for the lack of smoothness in arbitrary parts to derive the fundamental traction fields and the field equations. Moreover, we provide an alternative approach to derive the natural boundary conditions consistent with the postulates used to propose this theory.


%


\footnotesize

\bibliographystyle{unsrt}

\end{document}